\documentclass[preprint,preprintnumbers,amsmath,amssymb,superscriptaddress,nofootinbib,tightenlines]{revtex4} 

\usepackage{multirow}
\usepackage{graphicx}
\usepackage{dcolumn}
\usepackage{bm}

\begin{document}

\preprint{}


\title{Transition Form Factors and Mixing of Pseudoscalar Mesons from Anomaly Sum Rule}

\author{Yaroslav  Klopot}
\email{klopot@theor.jinr.ru}
\altaffiliation{on leave from Bogolyubov Institute for Theoretical Physics, 03680 Kiev, Ukraine}
\affiliation{Joint Institute for Nuclear Research, Dubna 141980, Russia}

\author{Armen  Oganesian}
\email{armen@itep.ru}
\affiliation{Joint Institute for Nuclear Research, Dubna 141980, Russia}
\affiliation{Institute of Theoretical and Experimental Physics, Moscow 117218, Russia}
\author{Oleg Teryaev}
\email{teryaev@theor.jinr.ru}
\affiliation{Joint Institute for Nuclear Research, Dubna 141980, Russia}


\begin{abstract}
Transition form factors of $\pi^0$, $\eta$, $\eta'$ mesons are investigated by means of the anomaly sum rule -- an exact nonperturbative relation which follows from the dispersive representation of the axial anomaly.  Considering the problem of contributions of operators originated from a non-(local) operator product expansion, we found that they are required by the available set of experimental data, including the most recent data from the Belle Collaboration (which, if taken alone, can be described without such contributions, although are compatible with them). In this approach we analyzed the experimental data on $\eta$ and $\eta'$ meson transition form factors and obtained the constraints on the decay constants and mixing parameters.

\begin{center}
\vskip 4cm

 \textit{\large We dedicate this paper to the memory of our friend and colleague Aleksander Bakulev.}

\end{center}

\end{abstract}

\maketitle

\section{Introduction}

The phenomenon of the axial anomaly \cite{Bell:1969ts,Adler:1969gk} plays an important role in nonperturbative QCD and
hadronic physics. The axial anomaly is known to govern the two-photon decays of the $\pi^0$, $\eta$, and $\eta'$ mesons and is usually considered for a case of real photons. However, the dispersive form of it \cite{Dolgov:1971ri}  can be considered for virtual photons also, \cite{Horejsi:1985qu,Horejsi:1994aj,Veretin:1994dn} leading to a number of interesting applications.

One of the consequences of the dispersive approach to the axial anomaly is a so-called anomaly sum rule (ASR) \cite{Horejsi:1994aj}. It gives, in particular,  a complementary way to describe the $\pi^0$ \cite{Klopot:2010ke} and the $\eta$, $\eta'$  \cite{Klopot:2011qq,Klopot:2011ai} transition form factors (later developed also in \cite{Melikhov:2012bg,Melikhov:2012qp}) at all $Q^2$, even beyond the QCD factorization. This is especially important in view  of the recent experimental studies of the $\gamma \gamma^* \to \pi^0 (\eta, \eta')$ transitions \cite{Aubert:2009mc,BABAR:2011ad,Uehara:2012ag}. In particular, the pion transition form factor, measured by the BABAR Collaboration \cite{Aubert:2009mc}, revealed  unexpectedly large values in the range of $Q^2=10$--$35$~GeV$^2$, resulting in an excess of the pQCD predicted  limit \cite{Lepage:1980fj} $Q^2 F_{\pi\gamma} \to \sqrt{2}f_\pi$, $f_\pi=0.1307$~GeV. This striking result attracted a lot of interest and motivated extensive theoretical investigations. As a result, the transition form factors  were (re)investigated  using the framework of  light cone sum rules \cite{Khodjamirian:2009ib,Mikhailov:2009sa,Bakulev:2011rp,Agaev:2010aq,Kroll:2010bf}, including  the flatlike modifications of the distribution amplitude \cite{Radyushkin:2009zg,Polyakov:2009je,Wu:2010zc}, the light-cone holography approach \cite{Brodsky:2011yv,Stoffers:2011xe}, and various model approaches, like a chiral quark model \cite{Dorokhov:2009dg} (see also \cite{Kotko:2009mb,Pham:2011zi,McKeen:2011aa,Lih:2012yu}), a vector meson dominance model, and its modifications \cite{Lichard:2010ap,Arriola:2010aq}. Some other approaches can be found in \cite{Kochelev:2009nz,Bystritskiy:2009bk,Noguera:2010fe,Roberts:2010rn,Guo:2011tq,Lucha:2011if,Czyz:2012nq,Noguera:2012aw,Dubnicka:2012zz,Masjuan:2012wy}. 


In this paper we extend and develop the anomaly sum rule approach \cite{Klopot:2010ke,Klopot:2011qq} to the transition form factors with a systematic account of the effects of  mixing of $\eta$, $\eta'$ mesons and quark-hadron duality. Also, we performed a new analysis of the anomaly sum rule for $\pi^0$ using available experimental data, including the most recent ones from the Belle Collaboration \cite{Uehara:2012ag}.

The paper is organized as follows. In Sec. II we give an overview of the anomaly sum rule approach and apply it to analyze  the pion (isovector channel of the ASR) and $\eta$, $\eta'$ (octet channel of the ASR) transition form factors. We pay a special attention to the seemingly controversial data from the BABAR \cite{Aubert:2009mc} and Belle \cite{Uehara:2012ag} collaborations. The analysis of different sets of data show that inclusion of the BABAR data requires a non-(local) operator product expansion (OPE) correction to the spectral density, while the  Belle data alone neither require nor exclude it. In the Sec. III  we develop and reformulate the description of mixing, which plays a special role for the $\eta$, $\eta'$ mesons, in a way which does not require the introduction of intermediate nonphysical states. The problem of compatibility of different mixing schemes is also discussed. In Sec. IV we perform a numerical analysis of the mixing parameters of the $\eta$-$\eta'$ system based on the obtained sum rule for the transition form factors. The possibility of the non-OPE correction to the spectral density in the octet channel is investigated as well.  The summary is presented in Sec V.

\section{Anomaly sum rule approach}

The  axial anomaly in QCD results in  a  nonvanishing divergence of axial current in the chiral limit. It is common to consider an octet of axial currents $J^{(a)}_{\mu 5}= \sum_{q} \bar{q} \gamma_5 \gamma_\mu \frac{\lambda^a}{\sqrt{2}} q$, ($a=1,..8$; the sum is over $u,d,s$ flavors; $\lambda^a$ are Gell-Mann matrices) and a singlet axial current $J^{(0)}_{\mu 5}=\frac{1}{\sqrt{3}}(\bar{u} \gamma_{\mu} \gamma_5 u + \bar{d} \gamma_{\mu} \gamma_5 d + \bar{s} \gamma_{\mu} \gamma_5 s)$. The singlet axial current  acquires both  electromagnetic and gluonic anomalous terms:  

\begin{align} \label{an-0}
\partial^\mu J_{\mu 5}^{(0)} =\frac{1}{\sqrt{3}}(m_u \overline{u}\gamma_5u+ m_d \overline{d}\gamma_5d + m_s\overline{s}\gamma_5s ) + 
 \frac{\alpha_{em}}{2\pi}C^{(0)}N_c  F\tilde{F}+\frac{\sqrt{3}\alpha_s}{4\pi} N_c G\widetilde{G},
\end{align}
where $F$ and $G$ are  electromagnetic and gluonic strength tensors, respectively; $\tilde{F}$ and $\tilde{G}$ are their duals; $N_c=3$ is the number of colors. On the contrary, diagonal components of the octet of axial currents, i.e.,  $J^{(3)}_{\mu 5}=\frac{1}{\sqrt{2}}(\bar{u} \gamma_{\mu} \gamma_5 u - \bar{d}
\gamma_{\mu} \gamma_5 d )$ and  $J^{(8)}_{\mu 5}=\frac{1}{\sqrt{6}}(\bar{u} \gamma_{\mu} \gamma_5 u + \bar{d}
\gamma_{\mu} \gamma_5 d - 2\bar{s} \gamma_{\mu} \gamma_5 s)$  acquire an electromagnetic anomalous term only:

\begin{align} \label{an-3}
\partial^\mu J_{\mu 5}^{(3)}& =\frac{1}{\sqrt{2}}(m_u \overline{u}\gamma_5u - m_d \overline{d}\gamma_5d ) + 
 \frac{\alpha_{em}}{2\pi}C^{(3)}N_c  F\tilde{F} ,\\
 \partial^\mu J_{\mu 5}^{(8)}& =\frac{1}{\sqrt{6}}(m_u \overline{u}\gamma_5u + m_d \overline{d}\gamma_5d -  2m_s \overline{s}\gamma_5s) +
 \frac{\alpha_{em}}{2\pi}C^{(8)}N_c  F\tilde{F}.\label{an-8}
\end{align}
The electromagnetic charge factors $C^{(a)}$  are

\begin{align} 
C^{(3)}&=\frac{1}{\sqrt{2}}(e_u^2-e_d^2)=\frac{1}{3\sqrt{2}},  \nonumber\\
C^{(8)}&=\frac{1}{\sqrt{6}}(e_u^2+e_d^2-2e_s^2)=\frac{1}{3\sqrt{6}}, \nonumber\\
C^{(0)}&=\frac{1}{\sqrt{3}}(e_u^2+e_d^2+e_s^2)=\frac{2}{3\sqrt{3}}.
\end{align}

In short, in what follows, we call $J_{\mu 5}^{(3)}$ and $J_{\mu 5}^{(8)}$ the isovector and octet current, respectively.


The calculation of the matrix elements of exact operator equations (\ref{an-3}) and  (\ref{an-8}), associated with the photon-meson transitions, leads to the triangle graph amplitude, composed of the axial current $J_{\alpha 5}$ with momentum $p=k+q$ and two vector currents  with momenta $k$ and $q$ [vector-vector-axial (VVA) amplitude]
\begin{equation} \label{VVA}
T_{\alpha \mu\nu}(k,q)=\int
d^4 x d^4 y e^{(ikx+iqy)} \langle 0|T\{ J_{\alpha 5}(0) J_\mu (x)
J_\nu(y) \}|0\rangle. 
\end{equation}
This amplitude can be decomposed \cite{Rosenberg:1962pp} (see also \cite{Eletsky:1982py,Radyushkin:1996tb}) as

\begin{align}
\label{eq1} \nonumber T_{\alpha \mu \nu} (k,q)  & =  F_{1} \;
\varepsilon_{\alpha \mu \nu \rho} k^{\rho} + F_{2} \;
\varepsilon_{\alpha \mu \nu \rho} q^{\rho}
\\ \nonumber
  & + \; \; F_{3} \; k_{\nu} \varepsilon_{\alpha \mu \rho \sigma}
k^{\rho} q^{\sigma} + F_{4} \; q_{\nu} \varepsilon_{\alpha \mu
\rho \sigma} k^{\rho}
q^{\sigma}\\
  & + \; \; F_{5} \; k_{\mu} \varepsilon_{\alpha \nu
\rho \sigma} k^{\rho} q^{\sigma} + F_{6} \; q_{\mu}
\varepsilon_{\alpha \nu \rho \sigma} k^{\rho} q^{\sigma},
\end{align}
where the coefficients $F_{j} = F_{j}(p^{2},k^{2},q^{2}; m^{2})$, $j = 1, \dots ,6$ are the corresponding Lorentz invariant
amplitudes constrained by current conservation and Bose symmetry. Note that the latter includes the interchange $\mu \leftrightarrow \nu$, $k \leftrightarrow q$ in the tensor structures and $k^2 \leftrightarrow q^2$ in the arguments of the scalar functions $F_{j}$.

In what follows, we consider  the case when  one of the photons is real ($k^2=0$) while the other is real or virtual ($Q^2=-q^2 \geq 0$).

For the isovector and octet currents,  using the dispersive treatment of the axial anomaly \cite{Dolgov:1971ri}, one can derive the anomaly sum rule (ASR) \cite{Horejsi:1994aj}:  

\begin{equation}\label{asr}
\int_{4m^{2}}^{\infty} A_{3}^{(a)}(s,Q^{2}; m^{2}) ds =
\frac{1}{2\pi}N_c C^{(a)} \;, a=3,8,
\end{equation}
where $A_{3} = \frac{1}{2}Im_{p^2} (F_3-F_6)$ and $m$ is a quark mass.

The ASR (\ref{asr}) has a remarkable property ---  both perturbative  and nonperturbative  corrections  to the integral  are absent \footnote{In the case of the singlet channel ($a=0$), this property is violated because of the gluonic anomaly}. The perturbative corrections are excluded because of the Adler-Bardeen theorem \cite{Adler:1969er}, while the nonperturbative corrections are also  absent, as is expected from 't Hooft's principle. 't Hooft's principle in its original form \cite{'tHooft:1980xb} implies that the anomalies of the fundamental fields are reproduced on the hadron level. In the dispersive approach this means the absence of the corrections to the dispersive sum rules.

Let us stress that the spectral density $A_{3}^{(a)}(s,Q^{2};m^{2})$ can have both perturbative  and nonperturbative corrections (however, the first-order correction $ \propto \alpha_s$ is  zero in the massless limit \cite{Jegerlehner:2005fs}), while the  integral $\int_{4m^{2}}^{\infty} A_{3}^{(a)}(s,Q^{2};m^{2}) ds$ equals exactly $\frac{1}{2\pi}N_c C^{(a)}$ .

Saturating the lhs of the three-point correlation function (\ref{VVA}) with the resonances and singling out their contributions to the ASR (\ref{asr}), we get (an infinite \cite{Klopot:2010ke}) sum of resonances with appropriate quantum numbers

\begin{equation} \label{qhd}
\pi \sum f_M^a F_{M\gamma} = \int_{4m^{2}}^{\infty} A_{3}^{(a)}(s,Q^{2};m^{2})
ds=\frac{1}{2\pi}N_c C^{(a)}.
\end{equation}
Here the projections of the axial currents $J^{(a)}_{5\alpha}$ onto
one-meson  states $M (=\pi^0, \eta, \eta')$  define the coupling (decay) constants $f^a_M$
\begin{equation} \label{def_f}
\langle 0|J^{(a)}_{\alpha 5}(0) |M(p)\rangle=
i p_\alpha f^a_M \;, 
\end{equation}
while the form factors $F_{M\gamma}$ of the transitions $\gamma\gamma^* \to M$  are defined by the matrix elements
	
\begin{equation} \int d^{4}x e^{ikx} \langle M(p)|T\{J_\mu (x) J_\nu(0)
\}|0\rangle = \epsilon_{\mu\nu\rho\sigma}k^\rho q^\sigma
F_{M\gamma} \;. 
\end{equation}
The relation (\ref{qhd}) expresses the global  duality between hadrons and quarks. 

\subsection{Isovector channel ($\pi^0$)}

For a case of the \emph{isovector channel}, the first contribution is given by $\pi^0$, while the higher contributions can be absorbed by the ``continuum'' contribution $\int_{s_0^{(3)}}^{\infty} A_{3}^{(3)}(s,Q^{2};m^{2})$, so the ASR (\ref{qhd}) takes the form

\begin{equation} \label{qhd3}
\pi f_{\pi}F_{\pi\gamma}(Q^2;m^2)  =\frac{1}{2\pi}N_c C^{(3)}-\int_{s_0^{(3)}}^{\infty} A_{3}^{(3)}(s,Q^{2};m^{2}) ds,
\end{equation}
where we assume for simplicity  $m=m_u=m_d$. 

The lower limit $s_0^{(3)}$ of the integral  we will refer to  as a ``continuum threshold'', bearing in mind that in a local quark-hadron duality hypothesis it means the interval of the duality of a pion. Also, it can be determined directly from the ASR, as we will later demonstrate.

The  contribution  to the spectral density  $ A_{3}^{(3)}(s,Q^{2};m^{2})$ for given flavor $q$ can be calculated from the VVA triangle diagram \cite{Horejsi:1994aj},

\begin{equation} \label{a3}
A_{3}^{(q)}(s,Q^{2};m_q^{2})=\frac{e_q^2}{2\pi}\frac{1}{(Q^2+s)^2}\left(Q^2R+2m_q^2 \ln \frac{1+R}{1-R}  \right),
\end{equation}
where $R(s,m_q^2)=\sqrt{1-\frac{4m_q^2}{s}}$ .

From (\ref{qhd3}), (\ref{a3}) a straightforward calculation gives an expression for the pion transition form factor,
\begin{align} \label{f3m}
F_{\pi\gamma}(Q^2;m^2)=\frac{1}{2\sqrt{2}\pi^2f_{\pi}}\frac{s_0^{(3)}}{s_0^{(3)}+Q^2}\times \nonumber\\ \Bigl [1-\frac{2m^2}{s_0^{(3)}}(\frac{2}{R_0+1}+\ln\frac{1+R_0}{1-R_0})\Bigr ],
\end{align}
where  $R_0=R(s_0^{(3)},m^2)$. This expression (to our best knowledge, for the first time) takes into account the contribution  of quark mass. 

Let us note that the quark mass term in (\ref{f3m}) (for $m \simeq 7$ MeV, and $s_0\simeq 0.7$ GeV$^2$) gives only $\simeq 0.15 \%$ contribution and can be neglected.

In the massless limit, the spectral density (\ref{a3}) is proportional to $\delta(s)$ at $Q^2=0$, so the continuum term in the ASR (\ref{qhd3}) goes to zero. This corresponds to the fact that contributions of axial states are zero at $Q^2=0$, and contributions of higher pseudoscalar states should be suppressed in order for the axial current to conserve in the chiral limit. 

Relying on the local quark-hadron duality hypothesis, the analysis of the two-point correlation function gives the value for the continuum threshold $s_0^{(3)}=0.75$ GeV$^2$ \cite{Shifman:1978by}.
Actually, $s_0^{(3)}$ can be determined directly from the  high-$Q^2$ asymptotic of ASR, where the QCD factorization predicts the value of the transition form factor $Q^2 F_{\pi\gamma}^{as}=\sqrt{2}f_{\pi}$ \cite{Lepage:1980fj}. The high-$Q^2$ limit of (\ref{f3m})  ($m=0$) immediately leads to  $s_0^{(3)}=4\pi^2f_{\pi}^2=0.67$ GeV$^2$. This expression, substituted  in (\ref{f3m}) with $m=0$, gives

\begin{equation} \label{f3bl}
F_{\pi\gamma}(Q^2;0)=\frac{1}{2\sqrt{2}\pi^2f_{\pi}}\frac{4\pi^2 f_{\pi}^2}{4\pi^2 f_{\pi}^2+Q^2},
\end{equation}
so it proves the Brodsky-Lepage interpolation formula for the pion transition form factor \cite{Brodsky:1981rp}, which was later confirmed by Radyushkin \cite{Radyushkin:1995pj}  in the approach of local quark-hadron duality.
Let us stress that  in this way we found  that it is a direct consequence of the anomaly sum rule (which is an exact nonperturbative QCD relation). 

It is interesting to note that by extending the expression for the pion transition form factor (\ref{f3m}) into the timelike region, one immediately gets a pole at $Q^2=s_0^{(3)}$ which is numerically close to the mass of a $\rho$ meson. This can be a kind of interplay between anomaly and vector dominance.

\begin{figure}
\includegraphics[scale=0.75]{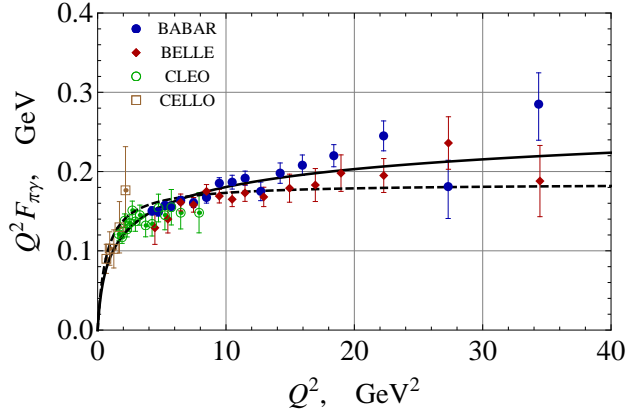}
\caption{\label{fig1} (color online). Pion transition form factor (multiplied by $Q^2$) with correction (Eq. (\ref{corr3F}), solid curve) and without correction (Eq. (\ref{f3bl}), dashed curve)  as a function of $Q^2$ compared with experimental data.}
\end{figure}

Currently the data from the CELLO \cite{Behrend:1990sr}, CLEO \cite{Gronberg:1997fj}, BABAR \cite{Aubert:2009mc} and Belle \cite{Uehara:2012ag} collaborations cover the region of $Q^2=0.7$--$35$ GeV$^2$ (see Fig. \ref{fig1}). While at $Q^2<10$ GeV$^2$ they are  consistent, at larger virtualities the BABAR  and a newly released Belle data are quite different. In this situation we will consider two sets of data:   CELLO+CLEO+Belle (I) and CELLO+CLEO+BABAR (II). 

When compared to the experimental data set (I), Eq. (\ref{f3bl}) gives a reasonable description consistent with the data  ($\chi^2/d.o.f.=1.01$, $d.o.f=35$, see the dashed line in Fig. \ref{fig1}). For the data set (II) the description is  worse ($\chi^2/d.o.f.=2.29$, $d.o.f.=37$). 
So, if the data of the BABAR Collaboration are correct, we come to a violation of the ASR-based expression for $F_{\pi\gamma}$ (\ref{f3bl}).

This means that the spectral density (\ref{a3}) must have a  substantial correction $\delta A_{3}$, which results in  corrections to the continuum $ \delta I_{cont}=\int_{s_0^{(3)}}^{\infty}  \delta A_3 ds$ and pion  $\delta I_{\pi}=\int_{4m^{2}}^{s_0^{(3)}} \delta A_{3} ds$  contributions. At the same time, as the \emph{full} integral remains constant, the corrections can be related: 

\begin{equation}
\delta I_{\pi} +\delta I_{cont}=0.
\end{equation}
It is important that the main terms of the continuum $I_{cont}$ and the pion $I_{\pi}$ contributions  have essentially different $Q^2$ behavior

\begin{align}
&I_{cont}=\int_{s_0^{(3)}}^{\infty}  A_3^{(3)}(s,Q^2) ds=\frac{1}{2\sqrt{2}\pi}\frac{Q^2}{s_0^{(3)}+Q^2},\\
&I_{\pi}=\int_{0}^{s_0^{(3)}}  A_3^{(3)}(s,Q^2) ds=\frac{1}{2\sqrt{2}\pi}\frac{s_0^{(3)}}{s_0^{(3)}+Q^2},
\end{align}
so the $1/Q^2$ power correction to the continuum contribution is of the order of the main term of the pion contribution.

Let us now discuss the sources of possible corrections  to the spectral densities which  in our approach  are the  counterparts of the nonlocal operator and higher twist corrections in the accurate pQCD fits \cite{Bakulev:2011rp}. Note that one-loop corrections to the spectral densities of all structures  in the VVA correlator in the massless case are zero, which can be easily deduced from the results of \cite{Jegerlehner:2005fs}.
If this nullification is due to conformal invariance \cite{Schreier:1971um,Gabadadze:1995ei},  one may expect the two-loop and higher  corrections to be nonzero due to the beta-function effects, which have recently been observed in the soft-photon approximation \cite{Mondejar:2012sz}. Nevertheless, these higher $\alpha_s$ corrections, as well as the local OPE-induced corrections, are small enough to produce an enhancement in the pion transition form factor shown by BABAR: $\delta F_{\pi\gamma} \sim log(Q^2)/Q^2$. Clearly, from dimensional arguments, such a term cannot  appear from the local OPE \cite{Klopot:2010ke}. Let us note also that even if some larger effective quark mass, instead of its current value, is taken in Eq. (\ref{f3m}), the mass term worsens  the experimental data description because of its negative sign.  Thus, to comply with the ASR, the correction should be of a non-(local) OPE origin, simulating the contribution of the operator of dimension 2. Among  the possible sources of such correction are nonlocal condensates, instantons, and short strings \cite{Chetyrkin:1998yr}.

Although the exact form of such a correction is not yet known, we can construct the simplest form of it relying on general requirements. Namely, the correction should vanish at $s_0^{(3)}\to \infty$ (the continuum contribution vanishes), at  $s_0^{(3)}\to 0$ (the full integral has no corrections), at  $Q^2\to \infty$ (the perturbative theory works at large $Q^2$)  and at $Q^2\to 0$ (the anomaly perfectly describes the pion decay width). Therefore, the correction satisfying those limits can be written as
  
\begin{equation}\label{corr3a}
\delta I_\pi = \frac{s_0^{(3)} Q^2}{(s_0^{(3)}+Q^2)^2}f(\frac{Q^2}{s_0^{(3)}}),
\end{equation}  
where $f$ is  a dimensionless function of $Q^2$, $s_0^{(3)}$ and some parameters.  Expecting the $\log (Q^2)/Q^2$ behavior, we can suggest the simplest (although not unique) form of the correction

\begin{equation}\label{corr3}
\delta I_\pi=\frac{1}{2\sqrt{2}\pi} \frac{\lambda s_0^{(3)} Q^2}{(s_0^{(3)}+Q^2)^2}(\ln{\frac{Q^2}{s_0^{(3)}}}+\sigma),
\end{equation}  
where $\lambda$ and $\sigma$ are dimensionless parameters. Then the pion transition form factor with this correction reads

\begin{align} \label{corr3F}
F_{\pi\gamma}(Q^2) = \frac{1}{\pi f_\pi} (I_\pi + \delta I_\pi)  = &  \nonumber 
\\ \frac{1}{2\sqrt{2}\pi^2f_{\pi}}\frac{s_0^{(3)}}{s_0^{(3)}+Q^2}\Bigl [1+\frac{\lambda Q^2}{s_0^{(3)}+Q^2}&(\ln{\frac{Q^2}{s_0^{(3)}}}+\sigma)\Bigr ].
\end{align}  
For the continuum threshold we will use $s_0^{(3)}=4\pi^2f_\pi^2=0.67$~GeV$^2$, which implies that at very high $Q^2$ the factorization restores. Nevertheless, if the factorization is violated at all $Q^2$, one can use a different value for $s_0^{(3)}$. Moreover, one can even consider the dependence  of $s_0^{(3)}$ on $Q^2$ \cite{Lucha:2011if}, which can lead to an effective change of $\sigma$. 
	
The fit of (\ref{corr3F}) to the data set (II) gives $\lambda=0.14,\;\sigma=-2.36$ with $\chi^2/d.o.f. =0.94 \;\; (d.o.f.=35)$. The plot of $Q^2F_{\pi\gamma}$  for these parameters is depicted in Fig. \ref{fig1} as a solid curve. Note that the proposed correction changes its sign at $Q^2 \simeq 8$~GeV$^2$, improving the description in both regions: at $Q^2 \lesssim 8$~GeV$^2$ and at $Q^2 \gtrsim 8$~GeV$^2$.
Also, $F_{\pi\gamma}$ (\ref{corr3F}) with these parameters $\lambda$,$\sigma$  (obtained from the fit to the data (II)) describes well also the data set (I) with $\chi^2/d.o.f. =0.84 \; (d.o.f.=35)$.

The summary of the fitting results for $F_{\pi\gamma}$ with (Eq. (\ref{corr3F})) and without (Eq. (\ref{f3bl}))  correction  for different sets of data is shown in Table I. One can see that the data sets involving the BABAR data require taking into account the correction. The data sets which do not involve the BABAR data may be described without such a correction, although the correction may  improve the description of the data.

\begin{table*} \label{table1}
\begin{center}
\begin{tabular}{|l|r|rrr|}
\hline
 & $\delta I_{\pi}=0:$~~~~~~ $\chi^2/d.o.f.$ & $\delta I_{\pi} \neq 0:$~~~ $\chi^2/d.o.f.$ & $\lambda$ & $\sigma$ \\
\hline
CELLO+CLEO+BABAR+Belle & 1.86 & 0.91 &  0.12 & -2.50 \\
CELLO+CLEO+Belle (set I) & 1.01 & 0.46 & 0.07 & -3.03 \\
CELLO+CLEO+BABAR (set II)& 2.29 & 0.94 & 0.14 & -2.36 \\
BABAR & 3.61 & 0.99 & 0.20 & -2.39 \\
Belle & 0.80 & 0.40 & 0.14 & -2.86 \\
\hline
\end{tabular}
\caption{\label{tab1} $\chi^2/d.o.f.$   obtained for Eq. (\ref{f3bl}) ($\delta I_{\pi}=0$) and the best fits of Eq. (\ref{corr3F}) ($\delta I_{\pi} \neq 0$) to different data sets.}
\end{center}
\vspace{-0.6cm}
\end{table*}

Let us emphasize that the correction (\ref{corr3})  requires a $\log Q^2$ term in $\delta A_3$ itself. This form of the correction  is different  from the one proposed in \cite{Melikhov:2012bg}  and could match it only if the prelogarithmic factor in (\ref{corr3}) did not depend on $s_0^{(3)}$. However, such a factor would violate  the above mentioned requirement of  nullification of the correction (in the limit $s_0^{(3)} \to 0$).

Also, there is a clear distinction with the natural emergence of the $\log Q^2$ term in the triangle amplitude (which was used for the description of the BABAR data), where the triangle amplitude itself is used as a model for the pion transition form factor \cite{Dorokhov:2009dg}. Such an approach applies the PCAC relation for the matrix elements involving large virtualities. In our opinion this procedure  is not justified to the same degree of rigor as for the soft processes. In our approach, the $\log Q^2$ term appears in the spectral density which is translated  to a transition form factor by an integral relation. Nevertheless,  we currently also cannot justify it by strict theoretical arguments.

\subsection{Octet channel ($\eta$, $\eta'$)}
In this subsection we consider the ASR for the octet channel, where, like for the isovector case, only the electromagnetic anomaly gives a contribution and the gluonic anomaly is absent. However, in comparing to the isovector channel, here we have some differences. 

First, due to significant mixing in  $\eta$--$\eta'$ system, the $\eta'$ meson  contributes to the octet channel. Since $\eta'$ decays into two real photons, it should be taken into account explicitly along with the $\eta$ meson. 

Second, the spectral density in the octet channel $A_3^{(8)}=\frac{1}{\sqrt{6}}(A_3^{(u)}+A_3^{(d)}-2A_3^{(s)})$ gets a more significant (in comparison to the isovector case)   mass contribution due to the strange quark.
Also there can be direct instanton contributions to the spectral density, which, however, should vanish in the massless limit as the singlet-octet transition is forbidden in the exact $SU(3)$ limit \footnote{The instanton contribution should also be absent in the integral of spectral density (\ref{asr}) for any mass due to 't Hooft's principle.}. This is in agreement with the consideration of the instanton contributions to the  two-point correlators \cite{Geshkenbein:1979vb}, where such contributions are $\propto m_s^2$  in the singlet-octet correlator and are absent in the octet-octet correlator.

In this paper we restrict ourselves to the leading approximation, where the quark mass corrections (both from triangle diagram (\ref{a3}) and direct instantons) to the spectral density $A_3^{(8)}$  are neglected.  Then treating the ASR in the same way as for the isovector channel gives

\begin{align} \label{asr8}
& f_{\eta}^8 F_{\eta\gamma}(Q^2) +f_{\eta'}^8F_{\eta'\gamma}(Q^2)= \frac{1}{2\sqrt{6}\pi^2}\frac{s_0^{(8)}}{s_0^{(8)}+Q^2},
\end{align}
where $f_{\eta}^{8}$, $f_{\eta'}^{8}$ are the decay constants defined in (\ref{def_f}), and $s_0^{(8)}$ is a continuum threshold in the octet channel.

As soon as this approximation provides a reasonable description of the experimental data (see Sec. IV), this may possibly indicate a partial cancellation of the instanton and mass effects in the VVA correlation function. 

Note also that the discrepancy with the two-photon decay width of the $\eta$ meson, considered in  \cite{Ioffe:2006ww} as a possible signal of instantons, is in fact eliminated when the mixing is taken into account \cite{Klopot:2010sm}, leading to specific values of $f_{\eta}^{8}$, $f_{\eta'}^{8}$.  This may also be interpreted as a partial absorption of the instanton contributions to the two-point correlation function by the  values of $f_{\eta}^{8}$, $f_{\eta'}^{8}$.

At the same time,  the reliable estimation of  $s_0^{(8)}$ from such a two-point correlator by the usual QCD sum rule method meets difficulties (see, e.g., discussion in \cite{Klopot:2011ai}). Fortunately, the ASR approach allows us to determine $s_0^{(8)}$ in the octet channel from the high-$Q^2$ asymptotic, just the same way as in the isovector channel. Generalization of the pion case gives the asymptotes for the $\eta$, $\eta'$ transition form factors \cite{Anisovich:1996hh,Feldmann:1997vc} ($M=\eta,\eta'$):

\begin{equation}
Q^2 F_{M\gamma}^{as}= 6(C^{(8)}f_M^8+ C^{(0)}f_M^0). 
\end{equation}

So,  the  $Q^2 \to \infty$ limit of the ASR (\ref{asr8}) gives
\begin{equation} \label{asr8inf}
s_0^{(8)}=4\pi^2((f_\eta^8)^2+(f_{\eta'}^8)^2+ 2\sqrt{2} [ f_\eta^8 f_{\eta}^0+ f_{\eta'}^8 f_{\eta'}^0]).
\end{equation}

In the  $Q^2=0$ limit the transition form factors are expressed in terms of the two-photon decay widths of mesons, so the ASR (\ref{asr8}) takes the form

\begin{equation} \label{asr8real}
f_{\eta}^8 F_{\eta\gamma}(0) +f_{\eta'}^8F_{\eta'\gamma}(0)=\frac{1}{2\sqrt{6}\pi^2},
\end{equation}
where
$$
F_{M\gamma}(0) =\sqrt{\frac{4\Gamma_{M\to \gamma\gamma}}{\pi \alpha^2 m^3_{M} }}.
$$

Solving Eqs. (\ref{asr8inf}), (\ref{asr8real}) with respect to $s_0^{(8)}$ and one of the decay constants $f^a_M$ and substituting them into general ASR (\ref{asr8}), we can relate the transition form factors $F_{\eta'\gamma}(Q^2)$ and the  decay constants $f^a_M$.

Actually, the four decay constants $f^a_M$ can be related based on a particular mixing scheme.
In \cite{Klopot:2011qq} the ASR was analyzed for several sets of  parameters and mixing schemes. 
The more general consideration of mixing schemes and extraction of the mixing parameters  are performed in the next sections.

\section{Mixing}

The problem of mixing in the $\eta$--$\eta'$ system is  usually  addressed either in the octet-singlet ($SU(3)$) or quark-flavor mixing scheme (see, e.g., \cite{Feldmann:1998vh,Feldmann:1999uf} and references therein).
Basically, meson mixing implies that the ``nondiagonal'' decay constants $f_{M}, M=\eta, \eta'$ in Eq. (\ref{def_f})  are nonzero. The $\pi^0$ and the isovector current can be decoupled from the $\eta$--$\eta'$ system and octet and singlet  currents because of a very small mixing.

Let us recall the common approach to the mixing, when  physical states are represented as a linear combination of states with definite $SU(3)_f$ quantum numbers or quark-flavor content. But, since these states do not have  definite masses, one cannot write the analogue of Eq. (\ref{def_f}) with these states instead of physical states $\eta$, $\eta'$. Indeed, if a state has a definite momentum $p^\mu$ it also has a definite mass $m^2=p_\mu p^\mu$. 

One can avoid this problem by formulating the mixing in terms of the fields $\phi_i$ related to the physical states $|i\rangle$ by the matrix elements $\langle 0|\phi_i|j \rangle=\delta_{ij}$.
 
It is well known that the field of the pion $\phi_{\pi}$ is defined from the divergence of the isovector component of the axial channel as a PCAC relation,
 \begin{equation}
 \partial_\mu J_{\mu5}^{(3)}=f_{\pi}^{(3)} m_{\pi}^2 \phi_{\pi}.
 \end{equation}

To consider the $\eta$ and $\eta'$ mixing, one can write down a straightforward generalization of this relation,
\begin{equation} \label{pcac0}
\mathbf{\partial_\mu J_{\mu5}=F M\Phi},
\end{equation}
where we introduced the matrix notations:

\begin{align} \label{defs}
&\mathbf{ J_{\mu 5}}\equiv
\begin{pmatrix}
 J_{\mu 5}^{\alpha}\\
 J_{\mu 5}^{\beta}
\end{pmatrix},
\mathbf{F}\equiv
\begin{pmatrix}
f_\eta^{\alpha} & f_{\eta'}^{\alpha} & f_G^{\alpha} & ... \\
f_\eta^{\beta} & f_{\eta'}^{\beta} & f_G^{\beta} & ...
\end{pmatrix}, 
\mathbf{\Phi}\equiv
\begin{pmatrix}
\phi_\eta\\
\phi_{\eta'}\\
\phi_G \\
\vdots
\end{pmatrix},\nonumber \\
&\mathbf{M}\equiv diag(m_\eta^2,m_{\eta'}^2,m_G^2, \cdots).
\end{align}

The  vector   $\mathbf{ J_{\mu 5}}$ in the lhs consists of the components of the axial current of definite  $SU(3)$ symmetry, a so-called octet-singlet basis ($\alpha =8$, $\beta =0$), or of the components of the axial current with the decoupled light and strange quark composition, a so-called quark-flavor basis ($\alpha = q$, $\beta=s$):
\begin{equation}
J_{\mu5}^q=\frac{1}{\sqrt 2}(\bar{u}\gamma_\mu \gamma_5 u + \bar{d}\gamma_\mu \gamma_5 d), \; J_{\mu5}^s=\bar{s}\gamma_\mu \gamma_5 s.
\end{equation}
The elements of matrix $\mathbf{F}$ are the meson decay constants defined in (\ref{def_f}). 
The vector  $\mathbf{\Phi}$ of physical fields contains the fields of $\eta$ and $\eta'$ mesons $\phi_\eta$ and $\phi_{\eta'}$ and the fields of higher mass states,  which we denote as $\phi_G, \hdots $

It is possible and common to get an additional model constraint for the matrix $\mathbf{F}$ which is fulfilled by applying the respective mixing scheme. Let us first introduce the new vector of fields $\mathbf{\widetilde \Phi}$  relating the $SU(3)$ symmetry property  or the quark-flavor contents of currents and meson fields. 
The first two  components of $\mathbf{\widetilde \Phi}$ are labeled with the same indices $\alpha$ and $\beta$ as the currents and correspond to the same symmetry ($\alpha =8, \beta =0$) or quark content ($\alpha = q, \beta=s$). The relation between $\mathbf{\Phi}$ and $\mathbf{\widetilde \Phi}$ is provided by the orthogonal  transformation $\mathbf{U}$

\begin{equation} \label{phi}
\mathbf{\widetilde{\Phi}=U\Phi}, \;\; \widetilde{\mathbf{\Phi}}\equiv
\begin{pmatrix}
\tilde \phi_{\alpha}\\
\tilde \phi_{\beta}\\
\tilde \phi_G \\
\vdots
\end{pmatrix}.
\end{equation}
In terms of these fields, Eq. (\ref{pcac0}) can be rewritten as

\begin{equation} \label{pcac2}
\mathbf{\partial_\mu J_{\mu5}=\widetilde{F} \widetilde{M}\widetilde {\Phi}},
\end{equation}
where $\mathbf{\widetilde{F}=FU}$, $\mathbf{\widetilde{M}=U^T M U}$.

In our notations the octet-singlet (quark-flavor) mixing scheme implies that  the matrix $\mathbf{\widetilde F}$ has a (rectangular) diagonal form in the respective octet-singlet (quark-flavor) basis,
\begin{equation}\label{Fdiag}
\mathbf{\widetilde F}=
\begin{pmatrix}
f_{\alpha} & 0 & 0 & ... \\
0 & f_{\beta} & 0 & ...
\end{pmatrix} .
\end{equation}

This relation can be obtained from the effective Lagrangian $\mathcal{L}$ which contains an interaction term $ \Delta \mathcal{L}_{int}=\frac{1}{2} \mathbf{\widetilde{\Phi}^T \widetilde{M} \widetilde{\Phi}} = \frac{1}{2}\sum\limits_{i,j} \tilde m_{ij}^2 \tilde \phi_i \tilde \phi_j$:
\begin{equation}
\partial_\mu J_{\mu5}^a =f_{a}\frac{\delta  \mathcal{L}}{\delta \tilde{\phi_a}} =f_{a} \sum\limits_{k} \tilde m_{ak}^2 \tilde{\phi}_k,\; a= \alpha,\beta.
\end{equation}

Note that from the requirement that matrix $\mathbf{FU}$ has a (rectangular) diagonal form (\ref{Fdiag}) immediately follows that $\mathbf{FF^T}$ is a diagonal matrix. So, imposing the mixing scheme is equivalent to imposing the constraint for the decay constants:
 \begin{equation} \label{ort-general}
  f_{\eta}^\alpha f_{\eta}^\beta+f_{\eta'}^\alpha f_{\eta'}^\beta+f_{G}^\alpha f_{G}^\beta+...=0.
 \end{equation}
Here the sum is over all physical meson states included in the vector $\mathbf{\Phi}$. 

If we restrict ourselves to consideration of the $\eta$ and $\eta'$ mesons only, then the decay constants form a $2\times 2$ matrix and in the octet-singlet and quark-flavor bases satisfy the respective diagonality constraints,

\begin{align} \label{ort-su3}
 f_{\eta}^8f_{\eta}^0+f_{\eta'}^8f_{\eta'}^0=0,\\
\label{ort-qf}
 f_{\eta}^qf_{\eta}^s+f_{\eta'}^q f_{\eta'}^s=0.
\end{align} 

For instance, in the case of the octet-singlet mixing scheme (\ref{ort-su3})  the matrix of decay constants can be expressed in terms of one mixing angle $\theta$ and two parameters $f_8$, $f_0$, forming the well-known one-angle mixing scheme:

\begin{equation} \label{1ang-80} \mathbf{F_{80}}=
 \left(\begin{array}{cc}  f_\eta^8 & f_{\eta'}^8 \\ f_\eta^0 & f_{\eta'}^0
\end{array}\right)=
 \left(\begin{array}{cc} f_8 \cos \theta & f_8\sin \theta \\
-f_0\sin \theta & f_0\cos \theta  \end{array}\right).
\end{equation}

Similarly, if the quark-flavor mixing scheme restriction (\ref{ort-qf}) is applied, then it is common to express the decay constants in terms of parameters $\phi$, $f_8$, $f_0$,
\begin{equation} \label{1ang-qs} \mathbf{F_{qs}}=
 \left(\begin{array}{cc}  f_\eta^q & f_{\eta'}^q \\ f_\eta^s & f_{\eta'}^s
\end{array}\right)=
 \left(\begin{array}{cc} f_q \cos \phi & f_q\sin \phi \\
-f_s\sin \phi & f_s\cos \phi  \end{array}\right).
\end{equation}

While either of the mixing schemes (octet-singlet or quark-flavor) is self-consistent, they are  incompatible \cite{Feldmann:1998vh,Klopot:2011ai}. Indeed, octet-singlet and quark-flavor  bases of axial currents are related by means of a rotation matrix:

\begin{equation}
\begin{pmatrix}
J_{\mu5}^8\\
J_{\mu5}^0\\
\end{pmatrix}=
\mathbf{V(\alpha)}
\begin{pmatrix}
J_{\mu5}^q\\
J_{\mu5}^s\\
\end{pmatrix},\; \mathbf{V(\alpha)}=
\begin{pmatrix}
\cos \alpha & -\sin \alpha \\
\sin \alpha & \cos \alpha \\
\end{pmatrix},
\end{equation}
where $\tan \alpha= \sqrt{2}$. Then, as follows from (\ref{pcac0}), the matrices of decay constants $\mathbf{F_{\alpha\beta}\equiv F}$ (\ref{defs}) in the octet-singlet (\ref{1ang-80}) and quark-flavor (\ref{1ang-qs}) bases are related as 
\begin{equation}\label{su3-qs}
\mathbf{F_{80}= V(\alpha) F_{qs}},
\end{equation}
and so  

\begin{equation}
\mathbf{ F_{80}F_{80}^T=V(\alpha)F_{qs}F_{qs}^T V(\alpha)^T}.
\end{equation}

We see that in the general case the decay constants cannot follow the octet-singlet and quark-flavor mixing scheme simultaneously (since the matrices $\mathbf{F_{80}F_{80}^T}$ and $\mathbf{F_{qs}F_{qs}^T}$ cannot be diagonal simultaneously).
The bases are compatible only if $f_8=f_0$ ($f_q=f_s$), i.e., in case of the exact $SU(3)_f$ symmetry.


Although the octet-singlet  mixing scheme is more convenient for the anomaly sum rule relations, which are exact in the cases of isovector and octet channels, there are arguments from  chiral perturbation theory against it \cite{Leutwyler:1997yr,HerreraSiklody:1996pm}  (see also \cite{Schechter:1992iz,Feldmann:1998vh}).

The mixing scheme-independent extraction of the decay constants from the experimental data can finally tell us which basis is more adequate for describing the mixing in the $\eta$--$\eta'$ system. This problem will be addressed in the next section.

\section{Octet channel: numerical analysis}

In Sec. II, as a consequence of the ASR in the octet channel, we obtained the relation between  transition form factors and decay constants of $\eta$ and $\eta'$ mesons. 
In this section we use this relation to analyze the decay constants in different  mixing schemes, described in Sec. III.

\begin{figure} 
\includegraphics[scale=0.7]{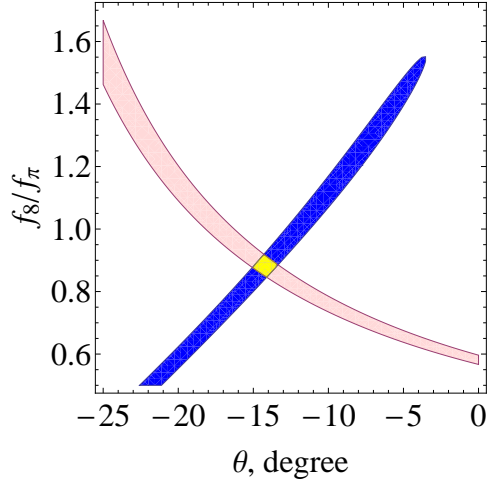}
\caption{\label{fig2}(color online). Octet-singlet mixing scheme parameters $f_8$, $\theta$. Dark blue region: constraint of  Eq. (\ref{asr8}) ($\chi^2/d.o.f.<1$). Light red region:  constraint of Eq. (\ref{asr8real}). (Experimental uncertainties are taken into account).}
\end{figure}

First, let us consider the \textit{octet-singlet mixing scheme}. In order to determine the mixing parameters of this scheme, we employ the mixing scheme constraint (\ref{ort-su3}) and  the ASR relations (\ref{asr8}), (\ref{asr8inf}), (\ref{asr8real}). In terms of the parameters  (\ref{1ang-80}), Eq. (\ref{asr8inf}) reads $s_0^{(8)}=4\pi^2f_8^2$. Then, the regions in $f_8, \theta$  parameter space which are constrained by the fit of Eq.  (\ref{asr8}) ($\chi^2/d.o.f.<1$) to the BABAR data \cite{BABAR:2011ad}  and by Eq. (\ref{asr8real}) (the experimental errors of $\Gamma_{\eta(\eta')\to 2\gamma}$ are taken into account) are shown in Fig. \ref{fig2}. The yellow intersection determines the parameters, which can be estimated as $f_8=(0.88\pm 0.04)f_\pi, \theta=-(14.2\pm 0.7)^\circ$. 

In order to determine the constant $f_0$ (which does not enter Eqs. (\ref{asr8}), (\ref{asr8inf}),(\ref{asr8real}) in the case of the octet-singlet mixing scheme), we need an additional constraint. 

As an additional constraint, it is convenient  to use the ratio of radiative decays of $J/\Psi$ for which the more cumbersome contribution of the gluonic anomaly is under control.
Indeed, according to Novikov \emph{et al.} \cite{Novikov:1979uy}, the radiative decays $J/\Psi \to \eta (\eta')\gamma$ are dominated by
the nonperturbative gluonic matrix elements, and the ratio of the
decay rates $R_{J/\Psi}=(\Gamma(J/\Psi)\to
\eta'\gamma)/(\Gamma(J/\Psi)\to \eta\gamma)$ is given by 

\begin{equation} \label{RJPsi} R_{J/\Psi}=\left|\frac{\langle0\mid
G\widetilde{G}\mid\eta'\rangle}{\langle0\mid
G\widetilde{G}\mid\eta\rangle}\right|^2\left(\frac{p_{\eta'}}{p_{\eta}}\right)^3,
\end{equation}
 where
$p_{\eta(\eta')}=M_{J/\Psi}(1-m^2_{\eta(\eta')}/M^2_{J/\Psi})/2$.

Taking matrix elements of the divergencies of the singlet (\ref{an-0}) and octet (\ref{an-8}) currents  between vacuum and $\eta (\eta')$ states and neglecting the $u$, $d$ quark masses and electromagnetic anomaly term,  the ratio (\ref{RJPsi}) can be expressed \cite{Ball:1995zv} in terms of the decay constants (\ref{def_f}) as follows:

\begin{equation} \label{RJPsi1} R_{J/\Psi}=\left(\frac{f_{\eta'}^8+\sqrt{2}f_{\eta'}^0}{f_{\eta}^8+\sqrt{2}f_{\eta}^0}\right)^2\left(\frac{m_{\eta'}}{m_{\eta}}\right)^4\left(\frac{p_{\eta'}}{p_{\eta}}\right)^3.
\end{equation}
The current experimental value of this ratio is $R_{J/\Psi}=4.67 \pm 0.15$ \cite{Beringer:1900zz}.

Employing this ratio for the octet-singlet mixing scheme and taking into account Eqs. (\ref{asr8}) and (\ref{asr8real}) one can determine  the singlet constant: $f_0=(0.81\pm 0.07)f_\pi$. So, the full set of constants of the octet-singlet scheme is

\begin{align}\label{par-os}
f_8&=(0.88\pm 0.04)f_\pi,\nonumber \\
f_0&=(0.81\pm 0.07)f_\pi, \nonumber\\
\theta&=-(14.2\pm 0.7)^\circ.
\end{align}

For the \textit{quark-flavor mixing scheme} we can perform a similar analysis, using the constraint of the scheme (\ref{ort-qf});  Eqs. (\ref{asr8inf}), (\ref{asr8real}),  (\ref{RJPsi1}); and fitting the ASR (\ref{asr8}) to the BABAR data \cite{BABAR:2011ad}. The decay constants of the quark-flavor basis $f_{\eta,\eta'}^{q,s}$ are expressed in terms of those of the octet-singlet basis $f_{\eta,\eta'}^{8,0}$ by means of Eq. (\ref{su3-qs}). In terms of the mixing parameters $f_q$, $f_s$, $\phi$ (\ref{1ang-qs}), Eqs. (\ref{asr8inf}), (\ref{RJPsi1}) read

\begin{align} \label{qf-s0}
s_0^{(8)}=&(4/3)\pi^2(5f_q^2-2f_s^2),\\ \label{qf-R} R_{J/\Psi}=&(\tan\phi)^2\left(\frac{m_{\eta'}}{m_{\eta}}\right)^4\left(\frac{p_{\eta'}}{p_{\eta}}\right)^3.
\end{align}

\begin{figure} 
\includegraphics[scale=0.7]{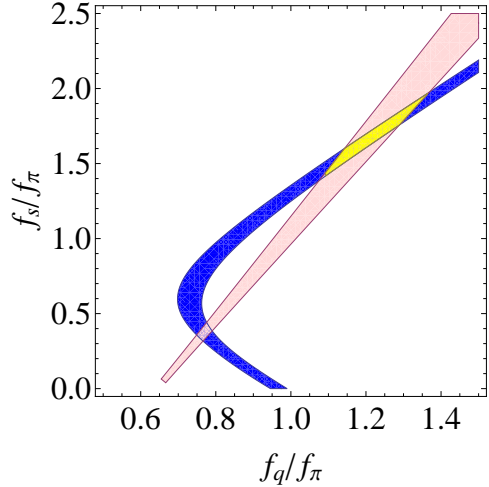}
\caption{\label{fig3} (color online). Quark-flavor mixing scheme parameters $f_q$, $f_s$. Dark blue region: constraint of  Eq. (\ref{asr8}) ($\chi^2/d.o.f.<1$). Light red region: constraint of Eq. (\ref{asr8real}). (Experimental uncertainties are taken into account).}
\end{figure}

Equation (\ref{qf-R})  determines the parameter $\phi=(38.1 \pm 0.5)^\circ$. Then the other two  parameters $f_s$, $f_q$ can be estimated from Eqs. (\ref{asr8real}) and (\ref{asr8}). The plot of the regions  constrained by these equations is shown in Fig. \ref{fig3}. The light red band indicates the constraint of Eq. (\ref{asr8real}) (experimental errors of $R_{J/\Psi}, \Gamma_{\eta(\eta')\to 2\gamma}$ are taken into account) and the dark blue band indicates the fit of Eq. (\ref{asr8}) to the BABAR data at the $\chi^2/d.o.f.<1$ level. One can observe two regions where both equations are compatible. We have chosen the physically motivated one, where $f_q,f_s>f_\pi$. So, the yellow intersection in Fig. \ref{fig3} determines the parameters  for the quark-flavor mixing scheme, which give the following ranges for them: 

\begin{align}\label{par-qf}
f_q&=(1.20\pm0.15)f_\pi,\nonumber \\
f_s&=(1.65\pm0.25)f_\pi, \nonumber\\
\phi&=(38.1\pm0.5)^\circ.
\end{align}
The obtained mixing parameters (\ref{par-qf}) are in agreement with those obtained in other approaches \cite{Feldmann:1998vh,Cao:1999fs,Escribano:2005qq}. 

\begin{figure}
\includegraphics[scale=0.8]{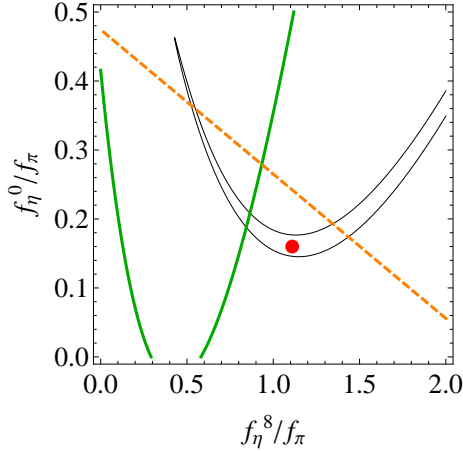}
\caption{\label{fig4} (color online). Independent (of mixing scheme) estimation of decay constants. The black thin curve is a $\chi^2/d.o.f.=1$ level; the red dot is a minimum of  $\chi^2$ of Eq. (\ref{asr8}). The green solid and orange dashed lines indicate the constraints of the octet-singlet (\ref{ort-su3}) and quark-flavor (\ref{ort-qf}) mixing schemes, respectively.}
\end{figure}

It is interesting also to get the \textit{mixing scheme independent} constraints on the decay constants. Equations (\ref{asr8real}), (\ref{RJPsi1}) allow us to exclude two of the four decay constants which enter the ASR (\ref{asr8}). The continuum threshold $s_0^{(8)}$ is excluded using (\ref{asr8inf}). In Fig. \ref{fig4}, the levels of the $\chi^2/d.o.f.$ function of Eq. (\ref{asr8}) in the space of constants $f_{\eta}^8$,  $f_{\eta}^0$ are shown (BABAR experimental data on $F_{\gamma \eta(\eta')} (Q^2)$ and mean values of $R_{J/\Psi}, \Gamma_{\eta(\eta')\to 2\gamma}$ are used). One can see, that the $\chi^2/d.o.f.<1$ requirement (black  curve) allows a rather wide range of the parameters. However, the minimum of $\chi^2$ is reached at the point $f_{\eta}^8=1.11 f_\pi$, $f_{\eta}^0=0.16 f_\pi$ ($\chi^2/d.o.f.=0.84$, indicated by a red dot). Then (\ref{asr8real}), (\ref{RJPsi1}) allow us to determine the other two constants:  $f_{\eta'}^8=-0.42 f_\pi$, $f_{\eta'}^0=1.04 f_\pi$. Therefore, the full set of decay constants of the mixing-scheme-independent  extraction is
\begin{equation}\label{par-indep}
 \left(\begin{array}{cc}  f_\eta^8 & f_{\eta'}^8 \\ f_\eta^0 & f_{\eta'}^0
\end{array}\right)=  \left(\begin{array}{cc}  1.11 & -0.42 \\ 0.16 & 1.04
\end{array}\right)f_\pi.
\end{equation}

The constraints of the octet-singlet (green solid curve) (\ref{ort-su3}) and quark-flavor (orange dashed curve) (\ref{ort-qf}) are also depicted in Fig. \ref{fig4}. We see that both considered mixing schemes are consistent with the scheme-independent analysis based on the ASR  at the level of $\chi^2/d.o.f.<1$, even if other experimental errors are not taken into account.  However, the least $\chi^2$ is reached in the region lying outside of both mixing scheme curves. Further improvement of the experimental data can clear up the question of the validity of different schemes and give more precise values of the  mixing parameters.

One can expect that the possible non-OPE  correction to the spectral density, discussed for the isovector channel, will  manifest in the same way in the octet channel also. Although the  BABAR data \cite{BABAR:2011ad} do not show such a sturdy growth of $Q^2F_{\eta\gamma}$ and $Q^2F_{\eta'\gamma}$  as they do for $Q^2F_{\pi\gamma}$, the octet combination of the transition form factors $Q^2(f_{\eta}^8 F_{\eta\gamma} +f_{\eta'}^8F_{\eta'\gamma})$ does reveal  a possible growth \footnote{The growth of the octet combination is due to the fact that  $Q^2F_{\eta\gamma}$ at $Q^2>20$ GeV$^2$ slightly tends to go up, and  $Q^2F_{\eta\gamma}$ tends to go down, while  $f_\eta^8$ is positive and $f_\eta^8$ is negative.}.

Expecting the similarity of the  correction to the spectral density in the isovector and octet channels,  we suppose that the correction in the octet channel has the same form as (\ref{corr3}),

\begin{equation}
\delta I_8= -\int_{s_0^{(8)}}^{\infty}  \delta A_3^{(8)} ds = \frac{1}{2\sqrt{6}\pi}\frac{\lambda s_0^{(8)} Q^2}{(s_0^{(8)}+Q^2)^2}(\ln{\frac{Q^2}{s_0^{(8)}}}+\sigma).
\end{equation}
This correction results in an additional term $\delta I_8/\pi$ in the rhs of (\ref{asr8}),

\begin{align} \label{asr8-corr}
f_{\eta}^8 F_{\eta\gamma}(Q^2) +f_{\eta'}^8F_{\eta'\gamma}(Q^2)=  \frac{1}{2\sqrt{6}\pi^2}\frac{s_0^{(8)}}{s_0^{(8)}+Q^2}+ \nonumber\\
\frac{1}{2\sqrt{6}\pi^2}\frac{\lambda s_0^{(8)} Q^2}{(s_0^{(8)}+Q^2)^2}(\ln{\frac{Q^2}{s_0^{(8)}}}+\sigma).
\end{align}

\begin{figure}
\includegraphics[scale=0.5]{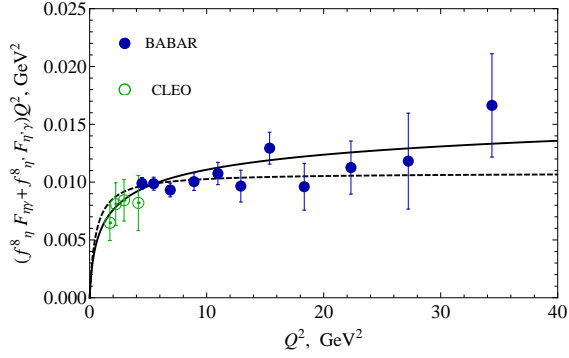}
\caption{\label{fig5a} (color online). ASR with correction  ((\ref{asr8-corr}), solid line) and without correction ((\ref{asr8}), dashed line) for the octet-singlet mixing scheme parameters (\ref{par-os}) compared with experimental data.}
\end{figure}

\begin{figure}
\includegraphics[scale=0.5]{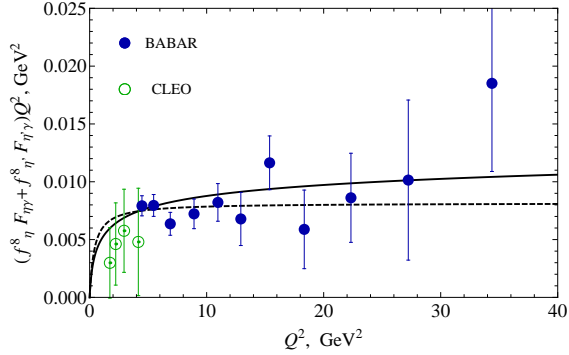}
\caption{\label{fig5b} (color online). ASR with correction  ((\ref{asr8-corr}), solid line) and without correction ((\ref{asr8}), dashed line) for the quark-flavor mixing scheme parameters (\ref{par-qf}) compared with experimental data.}
\end{figure}

\begin{figure}
\includegraphics[scale=0.5]{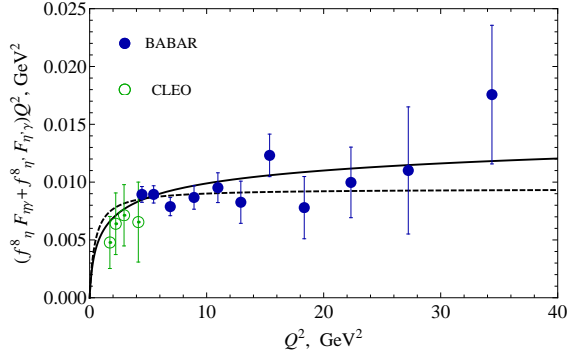}
\caption{\label{fig5c} (color online). ASR with correction  ((\ref{asr8-corr}), solid line) and without correction ((\ref{asr8}), dashed line)  for the mixing-scheme-independent parameters (\ref{par-indep}) compared with experimental data.}
\end{figure}

The plot of Eq. (\ref{asr8-corr}) (solid line) and Eq. (\ref{asr8}) (dashed line) for different mixing schemes are shown in Figs. \ref{fig5a}--\ref{fig5c}. The parameters $\lambda=0.14$,  $\sigma=-2.36$ are taken to be the same as those obtained for the pion case (data set II); the decay constants (central values) for different mixing schemes are employed from (\ref{par-os}), (\ref{par-qf}), and (\ref{par-indep}).  We see that the current precision of the experimental data on the $\eta$ and $\eta'$ transition form factors could accommodate the same correction as in the pion case, but does not require it.

Finally, let us make the following remark.
In this paper we developed the approach which does not rely on  introduction of the nonphysical states which have no definite masses. At  the same time,   
a hypothesis is widely discussed in the literature (see, e.g., \cite{Feldmann:1998yc,BABAR:2011ad}) that the transition form factor of the nonphysical state $|q \rangle \equiv \frac{1}{\sqrt{2}}(|\bar{u}u\rangle+|\bar{d}d\rangle)$ is related  to the pion form factor as $F_{q\gamma}(Q^2)=(5/3) F_{\pi\gamma}(Q^2)$ (where  the numerical factor  comes from the quark charges $(e_u^2+e_d^2)/(e_u^2-e_d^2)=5/3$).  
The states $|q\rangle$ and $|s\rangle \equiv|\bar{s}s\rangle$ are assumed to be  expressed in terms of the physical states $|\eta\rangle$, $|\eta'\rangle$ via the quark-flavor mixing scheme \footnote{Note that in our approach  the same relation exists only for the fields (see Eq. (\ref{phi})).},

\begin{equation}
|q\rangle =\cos \phi |\eta \rangle + \sin \phi |\eta' \rangle,\;\; |s\rangle =-\sin \phi |\eta \rangle + \sin \phi |\eta'\rangle.
\end{equation}
Then one can relate the form factors:

\begin{equation} \label{Fq}
\frac{5}{3}F_{\pi\gamma}=F_{\eta\gamma} \cos \phi + F_{\eta'\gamma} \sin \phi.
\end{equation}

Let us now try to incorporate  this hypothesis into our approach. For this purpose,  combining (\ref{asr8}) ($m=0$), (\ref{Fq}), and using (\ref{f3m}), (\ref{su3-qs})  we can express the $\eta, \eta'$ transition form factors in terms of the constants $f_q$, $f_s$,  $\phi$,

\begin{align}
F_{\eta\gamma}(Q^2)=&\frac{5}{12\pi^2f_s f_\pi}\frac{s_0^{(3)}(\sqrt{2}f_s\cos \phi -f_q\sin \phi)}{s_0^{(3)}+Q^2} + 
\frac{1}{4\pi^2f_s} \frac{s_0^{(8)}\sin\phi}{s_0^{(8)}+Q^2}, \label{feta}  \\
F_{\eta'\gamma}(Q^2)=&\frac{5}{12\pi^2f_sf_\pi} \frac{s_0^{(3)}(\sqrt{2}f_s\sin \phi +f_q\cos \phi)}{s_0^{(3)}+Q^2} - 
\frac{1}{4\pi^2f_s} \frac{s_0^{(8)}\cos\phi}{s_0^{(8)}+Q^2}, \label{fetap}
\end{align}
where $s_0^{(3)}=4\pi^2f_\pi^2$, $s_0^{(8)}=(4/3)\pi^2(5f_q^2-2f_s^2)$. 

\begin{figure}
\includegraphics[scale=0.55]{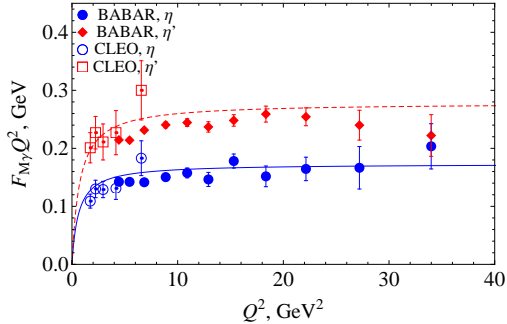}
\caption{\label{fig6} (color online). Combinations  $F_{\eta\gamma}Q^2$ (blue solid line) and $F_{\eta'\gamma}Q^2$ (red dashed line) from Eqs. (\ref{feta}) and (\ref{fetap}),  respectively, as functions of $Q^2$ compared with experimental data.}
\end{figure}

The plot of Eqs. (\ref{feta}), (\ref{fetap}) with constants from our analysis (\ref{par-qf}) $f_q=1.20f_\pi$, $f_s=1.65f_\pi$, $\phi=38.1^\circ$ in comparison with experimental data is shown in Fig. \ref{fig6}. One can observe a reasonably good agreement with the experimental data. 
For the decay constants of Ref. \cite{Feldmann:1998vh} one also gets a good description.

The agreement with the experimental data  may indicate that the effect of a strong anomaly for the $\frac{1}{\sqrt{2}}|\bar{u}u+\bar{d}d\rangle$ state  is small and the strong anomaly predominantly  appears in the $\bar{s}s$ channel. This statement can be rigorously checked by means of the anomaly sum rule for the singlet channel, which we postpone to future work.

\section{Summary and discussion}

The exact anomaly sum rule allows us to derive an expression  for the pion transition form factor at arbitrary $Q^2$, giving the proof for the Brodsky-Lepage interpolation formula. At $Q^2=0$ it is related to the pion decay width, while at large $Q^2$, basing on the factorization approach, it allows us to define the pion interval of duality which numerically  appears to be close to the value defined from the two-point correlator sum rule analysis.
However, the proposed approach may be applied even when the factorization is broken. It was exactly the situation supported by BABAR data requiring a small nonperturbative correction \cite{Klopot:2010ke} to the continuum spectral density. Having dimension 2, it cannot appear in (a local) OPE and should be attributed  to, say, instantons or short strings.

In this work we included in our analysis the recent Belle Collaboration data. The main conclusions are the following. Although the Belle data themselves may be described without the mentioned correction, they do not also exclude its possibility. Unless the BABAR data are disproved, the need for the correction remains. This is supported by the Table 1, where the fits for various combinations of the data are shown.

The search for the discussed correction can be performed also by means of lattice simulations, which already provided evidence (see, e.g., \cite{Blossier:2010vt} and references therein) for non-OPE vacuum condensates. In our case one may study the three-point VVA correlator on the lattice in a way  similar to \cite{Owen:2012ej}. To be sensitive to the discussed correction, one should consider moderately large momentum transfer in one of the vector channels. 

The corrections in the VVA correlator can be also studied analytically by generalization of the approach used in \cite{Chetyrkin:1998yr} to the case of the three-point correlation function.
Some indications of dimension 2 corrections can also be obtained  by the refined analysis \cite{inprep} of $e^+ e^-$ annihilation data.

In the generalization of our approach to the $\eta$ and $\eta'$ mesons, the data may be described without such a correction.  However, the possibility of the correction, similar to that discussed for the pion case, is not excluded by the current experimental data and is even supported by the slight growth of the octet combination of transition form factors.
So we can conclude that the correction to the spectral density, first introduced in \cite{Klopot:2010ke}, seems to be universal for both isovector and octet channels. This conclusion is in an agreement with a recent discussion in \cite{Melikhov:2012qp}.

The mixing plays a special role in the octet channel. There are two mixing models on the market now: the octet-singlet and quark-flavor mixing schemes.  We reformulate these models without introducing the nonphysical states with indefinite masses. Each scheme implies a certain constraint for the meson decay constants  (\ref{ort-su3}), (\ref{ort-qf}).
Both mixing models are compatible only in the exact $SU(3)_f$ limit.

Using the data on the transition form factors of the $\eta$, $\eta'$ mesons, the ASR allows us to extract the set of decay constants in the octet-single and quark-flavor schemes, as well as in the mixing-scheme-independent way, if we add an additional constraint of the ratio of radiation decays of the $J/\psi$ meson (\ref{RJPsi1}). It is shown that the current data precision permits both the octet-singlet and quark-flavor mixing schemes. Future improvements to experimental data on  transition form factors of $\eta$, $\eta'$ mesons and the ratio $R_{J/\psi}$, expected from the Belle and BES-III collaborations, can determine which scheme is more suitable for the description of mixing in the $\eta$--$\eta'$ system.

\begin{acknowledgments}

We thank  \boxed{A.~Bakulev}, S.~Brodsky, S.~Eidelman, B.~Ioffe, F.~Jegerlehner, A. Kataev, D. Melikhov, S.~Mikhailov,  A.~Pimikov, M.~Polyakov,  N.~Stefanis, and S.~Uehara for  discussions and useful comments. Y.K. thanks  Y.~Okada for warm hospitality at KEK, where a part of this work was completed. A.O. and O.T. thank N.~Stefanis and M.~Polyakov for  warm hospitality during their stay at ITP, Ruhr-University Bochum. This work is supported in part by RFBR, research projects 12-02-00613a, 12-02-00284a, and by the Heisenberg-Landau program (JINR).
\end{acknowledgments}

\end{document}